\documentclass[pdflatex,sn-nature]{sn-jnl}

\usepackage{graphicx}%
\usepackage{multirow}%
\usepackage{amsmath,amssymb,amsfonts}%
\usepackage{amsthm}%
\usepackage{bm}
\usepackage{mathrsfs}%
\usepackage[title]{appendix}%
\usepackage{xcolor}%
\usepackage{textcomp}%
\usepackage{manyfoot}%
\usepackage{booktabs}%
\usepackage{algorithm}%
\usepackage{algorithmicx}%
\usepackage{algpseudocode}%

\usepackage{listings}%

\usepackage{upgreek}
\usepackage{gensymb}

\begin{document}

%\title[Article Title]{Pockels Coefficient for ScAlN at various Sc concentrations : Ab-initio Simulation and Experiments}

\title[Article Title]{Unveiling the Pockels Coefficient of Ferroelectric Nitride ScAlN}


\author[1]{\fnm{Guangcanlan} \sur{Yang}}

\author[2]{\fnm{Haochen} \sur{Wang}}

\author[3]{\fnm{Sai} \sur{Mu}}

\author[1]{\fnm{Hao} \sur{Xie}}
\author[4,5]{\fnm{Tyler} \sur{Wang}}
\author[1]{\fnm{Chengxing} \sur{He}}

\author[1]{\fnm{Mohan} \sur{Shen}}
\author[1,5] {\fnm{Mengxia} \sur{Liu}}
\author[2]{\fnm{Chris G.} \sur{Van de Walle}}
\author*[1]{
\fnm{Hong X.} \sur{Tang}}\email{hong.tang@yale.edu}

\affil[1]{\orgdiv{Department of Electrical Engineering}, \orgname{Yale University}, \orgaddress{\city{New Haven}, \state{CT}, \postcode{06511}, \country{USA}}}

\affil[2]
{\orgdiv{Materials Department}, \orgname{University of California Santa Barbara}, \orgaddress{\city{Santa Barbara}, \state{CA}, \postcode{93106}, \country{USA}}}

\affil[3]{
\orgdiv{SmartState Center for Experimental Nanoscale Physics, Department of Physics and Astronomy}, \orgname{University of South Carolina}, \orgaddress{{\city{Columbia}, \state{SC}, \postcode{29208}, \country{USA}}}
}

\affil[4]{\orgdiv{Department of Physics}, \orgname{Yale University}, \orgaddress{\city{New Haven}, \state{CT}, \postcode{06511}, \country{USA}}}

\affil[5]{\orgdiv{Energy Sciences Institute}, \orgname{Yale University}, \orgaddress{\city{West Haven}, \state{CT}, \postcode{06516}, \country{USA}}}


\abstract{
Nitride ferroelectrics have recently emerged as promising alternatives to oxide ferroelectrics due to their compatibility with mainstream semiconductor processing. ScAlN, in particular, has exhibited remarkable piezoelectric coupling strength ($K^2$) comparable to that of lithium niobate, making it a valuable choice for RF filters in wireless communications. Recently, ScAlN has sparked interest in its use for nanophotonic devices, chiefly due to its large bandgap facilitating operation in blue wavelengths coupled with promises of enhanced nonlinear optical properties such as a large second-order susceptibility ($\chi^{(2)}$). It is still an open question whether ScAlN can outperform oxide ferroelectrics concerning the Pockels effect---an electro-optic coupling extensively utilized in optical communications devices. In this paper, we present a comprehensive theoretical analysis and experimental demonstration of ScAlN's Pockels effect. Our findings reveal that the electro-optic coupling of ScAlN, despite being weak at low Sc concentration, may be significantly enhanced and exceed LiNbO$_3$ at high levels of Sc doping, which points the direction of continued research efforts to unlock the full potential of ScAlN.
}

%%================================%%
%% Sample for structured abstract %%
%%================================%%

% \abstract{\textbf{Purpose:} The abstract serves both as a general introduction to the topic and as a brief, non-technical summary of the main results and their implications. The abstract must not include subheadings (unless expressly permitted in the journal's Instructions to Authors), equations or citations. As a guide the abstract should not exceed 200 words. Most journals do not set a hard limit however authors are advised to check the author instructions for the journal they are submitting to.

%\keywords{keyword1, Keyword2, Keyword3, Keyword4}



\maketitle

\section*{Introduction}
Over the past decade, the interest in ferroelectric materials has surged owing to their possibilities for diverse applications in memory cells, actuators, nonlinear optical devices, and more\cite{2020-mikolajick-ferromemory,2022-zetian-mbememory,2000-muralt-ferroelectricactuator,2021-kreutzer-actuator,1972-kielich-ferroelectricNLO,2020-yifan-LNNLO}.
While extensive research has been performed on oxide ferroelectrics such as lithium niobate (LN) and barium titanate (BTO), their substrate production processes remain challenging to be adopted into cost-effective, large-scale industrial production\cite{2012-hui-LNOIsmartcut,2017-kristy-btogrowth}. 
Nitride ferroelectric materials, on the other hand, are increasingly gaining recognition as viable CMOS-compatible alternatives\cite{2015-wong-cmosgan,2021-valerie-chi2,2023-kim-cmosAlScN}. Among them, scandium-alloyed aluminum nitride (ScAlN) has emerged as a prominent candidate, largely due to its crystallizing in the same wurtzite structure as unalloyed AlN\cite{2002-noboru-abiScAlN} and due to its enhanced piezoelectric coefficient% at Sc doping levels near 43$\%$
\cite{2021-ambacher-scalnproperty,2009-morito-43scalnpiezo}, making ScAlN an excellent material for micro-electromechanical systems (MEMS)\cite{2009-morito-43scalnpiezo,2017-wang-mems,2017-colombo-mems}. 
Some prior studies have also confirmed that the second-order nonlinear susceptibility ($\chi^{(2)}$) of ScAlN can surpass that of LN in both sputter-deposited and molecular beam epitaxy (MBE)-grown thin films\cite{2021-valerie-chi2,2023-jiangnan-chi2}. 
Additionally, ScAlN possesses a large direct bandgap exceeding 4.5\,eV\cite{2019-martina-bandgap}, thus permitting operation in the deep-blue wavelength range. 
All combined, these properties show great potential and prospects of ScAlN being a superior material for both linear and nonlinear photonic applications.

One of the most fundamental applications of nonlinear optical materials is electro-optic (EO) modulation, which has served a broad spectrum of applications including telecommunications, quantum optics, optical computing, and biomedical imaging\cite{2004-jeffrey-EOMcomm,2024-mohan-LNSEOM,2019-luo-EOMqo,2021-kumar-EOMopticcomp,2018-zhen-EOMimaging}. Creating effective EOMs is largely dependent on finding nonlinear materials that simultaneously posses high Pockels coefficients and low optical loss.
As such, the pursuit for electro-optic materials amenable to mass commercial production remains an open issue, for which ScAlN offers a promising solution. 
Nevertheless, to the best of our knowledge, the Pockels coefficient of ScAlN has not been systematically investigated. The lack of bulk single crystals make direct measurements particularly difficult. Furthermore, realizing ScAlN's promising materials properties in effective photonic devices has proven challenging, largely due to difficulty in fabricating integrated ScAlN photonic circuits on suitable low refractive index substrates. Previous device fabrication attempts have also suffered from significant optical loss in the ScAlN photonic waveguide due to a rough sidewall after etching\cite{2020-zhu-scalnphotonic}. Empirically, varying deposition conditions also lead to large differences in the final ScAlN film qualities, making systematic optimization of integrated photonic devices difficult. Despite these experimental challenges, ab-initio density-functional theory (DFT) calculations have emerged as a powerful tool for predicting physical properties of new materials. This method has successfully predicted the existence of wurtzite-ScAlN\cite{2002-noboru-abiScAlN}, and estimated the piezoelectric properties of ScAlN in good agreement with experimental results\cite{2010-ferenc-piezoorigin,2021-haochen-piezo}, validating DFT as a viable means of studying this important material family.  

Here, we first characterize the material properties of sputter-deposited ScAlN thin film with Sc concentrations of 0\%, 10\% and 30\%. DFT calculations show good agreement with the measured lattice parameters. A low-loss ScAlN-on-insulator photonic platform is then realized through transferring thin film ScAlN from commercially available wafers onto an insulator-on-Si substrate. The Pockels coefficients of ScAlN are experimentally measured with fabricated electrically tunable microring resonators, the results of which are further verified by DFT calculations. We show that the EO response of ScAlN with low Sc concentration is not as enhanced as previously hoped, while it is still very promising to achieve intense EO response at high Sc concentrations according to our calculations. Interestingly, the $r_{13}$ Pockels coefficient displays an inversion of sign between the 10\% and 30\% films. 
Using DFT, we can attribute this inversion to the ionic contribution in the response function: the Raman susceptibility of the A1 mode that dominates this term shifts and changes sign as Sc is added to the alloy.

\section*{Results}
\subsection*{ScAlN: material characterization and first-principles calculations}\label{sec2}

The deposition of high-quality ScAlN thin films has been an important topic within materials science. To date, two primary methods are employed for the ScAlN growth: magnetron sputtering and molecular-beam epitaxy (MBE). These techniques yield poly-crystalline and single-crystalline films respectively. MBE-grown nitride ferroelectrics generally exhibit superior optical performance compared to their sputtered counterparts\cite{2012-chi-aln,zetian-2019-highQaln}. However, the necessity for specific lattice-matched substrates constrains the Sc concentration to approximately 11-18$\%$ when grown atop GaN buffers\cite{2022-zetian-mbememory,DJ-2023-mbeLattice}. In contrast, sputter-deposition offers a more flexible choice of substrate and allows a greater range of possible Sc concentrations\cite{2021-ambacher-scalnproperty}. 
Our ScAlN wafers are deposited through a dual target AC magnetron sputtering by a commercial foundry (Advanced Modular Systems, Inc.)\cite{2023-yury-cosputter}. 
For better seeding of columnar ScAlN growth, the Si substrates are first coated with AlN followed by Mo. Two target Sc concentrations (10$\%$ and 30$\%$) are prepared, with a respective thickness of 1,000 and 600\,nm. For comparison, a 400\,nm-thick reference AlN sample (0\% Sc) is also prepared by sputter-deposition on an insulator substrate.

\begin{figure}[t]
    \centering
    \includegraphics[width=\textwidth]{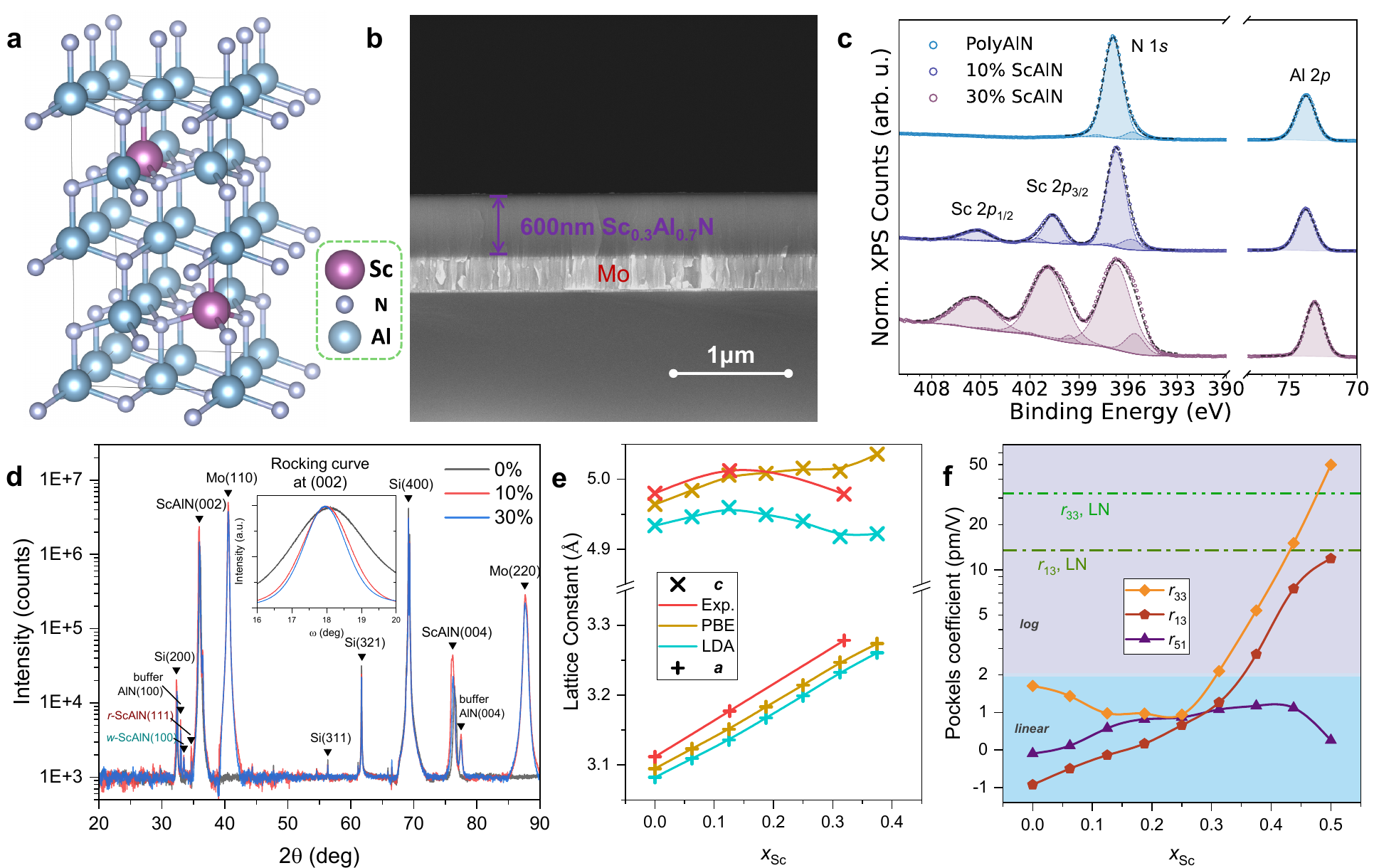}
    \caption{Characterization and first-principles study of ScAlN. (a) 32-atom unit cell of wurtzite ScAlN with Sc concentration of 12.5$\%$. (b) Cross-sectional SEM imaging of 600\,nm-thick Sc$_{0.3}$Al$_{0.7}$N, presenting the poly-crystalline nature of sputter-deposited ScAlN thin films. (c) XPS analysis of ScAlN films and the reference poly-AlN sample, yielding an analytical Sc concentration of $12.6\pm1.2\%$ and $31.9\pm3.2\%$ for the two Sc-alloyed samples.  (d) Out-of-plane XRD containing characteristic peaks for ScAlN, Si, and Mo. Inset: Rocking curve measurement showing FWHMs $\leq1.6\degree$ for both ScAlN samples. AlN sample sputtered on insulator substrate exhibits a larger $2.6\degree$ FWHM. (e) Comparison between measured (Exp.) and DFT-calculated lattice parameters at various Sc concentrations. (f) Calculated Pockels coefficients $r_{13}, r_{33}$, and $r_{51}$. Notably, $r_{33}$ has the potential to exceed that of LN at high levels of Sc alloying. A log/linear scale is adopted for the $y$-axis, marked by purple and blue background colors respectively.}
    \label{fig1}
\end{figure}

Scanning electron microscopy (SEM) depicted in Fig.\,\ref{fig1}b reveals that a columnar structure presents in the sputter-deposited samples, arising from the poly-crystalline nature of the films. Each column exhibits a diameter on the order of tens of nanometers, representing a single crystallite characterized by its unique in-plane lattice orientation. 

Core-level X-ray photoelectron spectroscopy (XPS) is carried out to study the chemical state and concentration of each element in ScAlN.  Fig.\,\ref{fig1}c shows that the Al\,2$p$ and dominant N\,1$s$ core-spectrum of poly-AlN appears at 73.4\,eV and 396.9\,eV respectively, in agreement with previous reports \cite{2008-rosenberger-XPSScAlN}. Upon the introduction of Sc, no noticeable splitting and negligible shifting of the Al spectrum was observed, suggesting weak Al-Sc interactions and that Sc is well integrated into the lattice. In the 10\% Sc-alloyed sample, additional Sc\,2$p$ core-level peaks were observed at around 405.4 and 400.6\,eV. At higher concentrations, these peaks shift to higher binding energies, reflecting stronger Sc interactions in the more strongly alloyed system. After peak fitting, we analytically obtain the cation atomic percentage to be approximately $12.6\pm1.2\%$ and $31.9\pm3.2\%$ Sc for the two samples. Notably, the small side peak contributions to the spectrum can be attributed to certain defects, with more detailed analysis available in supplementary information (SI).

%To further assess the lattice structure and crystallinity information of the ScAlN samples, 
Figure~\ref{fig1}d overlays the X-ray diffraction (XRD) patterns of all samples, with identified peaks of ScAlN, Si, and Mo duly labeled. The distinct (Sc)AlN\,(002) peaks, appearing at approximately $36\degree$, indicate a predominance of the $c$-axis orientation perpendicular to the plane of the substrate. Additionally, the ScAlN films exhibit extra diffraction peaks around $33.5\degree$ and $34.7\degree$, potentially corresponding to the first-order diffraction from the wurtzite ScAlN’s (100)-plane ($a$-plane) and the rock-salt ScAlN’s (111)-plane, respectively. The occurrence of these peaks corresponds with the presence of abnormally oriented grains (AOG) on ScAlN surface (SI Fig.\,S1)\cite{2019-sandu-aog}. Furthermore, the rocking curve analysis of the (002) peak, as shown in the inset of Fig.\,\ref{fig1}d, reveals a FWHM of $1.6\degree$ for the 10\% and $1.4\degree$ for the 30\% ScAlN films, both surpassing the quality of other sputter-deposited films reported in the literature \cite{2021-tsai-rockcurve,2022-jingxiang-rockcurve}. In contrast, the AlN sample exhibits a FWHM of $2.6\degree$, implying that direct sputter deposition on an insulator substrate could significantly compromise film quality due to lattice mismatch.

To qualitatively understand how Sc concentration impacts the Pockels effect of ScAlN, we perform first-principles calculations using 32-atom supercells as illustrated in Fig.\,\ref{fig1}a. 
The corresponding Sc concentrations range from 0\% to 50\% with 6.25\% intervals.
All our calculations are based on density functional theory using the \texttt{ABINIT} software package\cite{gonze2005brief}.
The optimized norm-conserving Vanderbilt pseudopotentials (ONCVPSP) used in the calculations are generated in PseudoDojo\cite{hamann2013optimized, van2018pseudodojo}.  
Two functionals, local density approximation (LDA)\cite{ceperley1980ground, perdew1981self}, and generalized gradient approximation (GGA) by Perdew, Burke, and Ernzerhof (PBE)\cite{PBE}, are employed to account for exchange and correlation in different calculations.
At each Sc concentration, we generate at least five inequivalent configurations with atoms randomly placed and calculate the average of various properties. We use a plane-wave cutoff energy of 45 hartree and a $3 \times 3 \times 2$ Monkhorst-Pack k-point grid \cite{monkhorst1976special}. All structures are relaxed until the interatomic forces are smaller than $5 \times 10^{-5}$ hartree/bohr. 

The lattice parameters observed within the relaxed structures are further compared with experimental data obtained from XRD measurements. As indicated in Fig.\,\ref{fig1}e, both the experimental and computational data indicate a linear relationship between the $\bm{a}$ constant and Sc concentration. This observation is consistent with findings reported in Ref.\,\cite{2021-ambacher-scalnproperty} and \cite{2021-haochen-piezo}. Meanwhile, the empirical $\bm{c}$ constant exhibits an initial increase at 10$\%$ Sc concentration, followed by a decrease at 30$\%$. This behavior mirrors the trend predicted by LDA calculations. The PBE results, on the other hand,  show a monotonic increase with Sc concentration. This suggests that LDA may provide a more accurate prediction of ScAlN's physical properties.

Based on the LDA results, we employ density functional perturbation theory \cite{veithen2005nonlinear} to calculate the Pockels coefficients for ScAlN. We adopt the Voigt notation, in which EO coefficients $r_{ijk}$ are contracted to a 6$\times$3 matrix. 
In an ideal wurtzite crystal with 6\textit{mm} symmetry the only non-vanishing elements of the matrix should be $r_{13}=r_{23}, r_{33},$ and $r_{51}=r_{42}$. 
However, due to the symmetry lowering caused by alloying Sc into AlN,
other elements in the EO tensor, such as $r_{11}$ and $r_{62}$, may also be nonzero.
We observe that after configurational averaging over many supercells, the $r_{11}$ and $r_{62}$ values are close to zero. 

As depicted in Fig. \ref{fig1}f, we present the calculated EO coefficients $r_{13}$, $r_{33}$ and $r_{51}$ as a function of Sc concentration. 
The $r_{33}$ coefficient initially decreases and then rises at a Sc concentration of 25\%. Meanwhile, $r_{13}$ monotonically increases across the entire range of Sc concentrations, crossing zero value at approximately 16\%.
At high Sc concentrations (43\% and 50\%), both $r_{13}$ and $r_{33}$ experience significant enhancements, approaching or even exceeding those observed in lithium niobate \cite{2020-yifan-LNNLO}.
These enhancements are primarily attributed to the piezoelectric contributions to the EO coefficients and are influenced by a short-range order that can be present in ScAlN alloys at high Sc concentrations \cite{wang2024-arxiv}.
Consequently, ScAlN with higher Sc fraction holds great promise for  EO applications that take advantage of $r_{13}$ and $r_{33}$ coefficients.
In contrast, $r_{51}$ does not exhibit enhancements at high Sc concentrations, remaining less than 2~pm/V at all calculated concentrations.

\subsection*{ScAlN-on-insulator-on-silicon photonics platform }\label{sec3}
Accurate determination of Pockels coefficients in thin film materials necessitates the construction of an appropriate photonic device. Traditionally, the fabrication of integrated ScAlN photonic circuits has been hindered by challenges associated with material preparation on suitable low refractive index substrates. Direct sputter-deposition on insulators can lead to a degradation of film quality, as is indicated by the broadened XRD rocking curve. While depositing ScAlN films on Mo ensures a higher quality film, it unavoidably leads to large metal absorption at optical wavelengths, thus preventing these wafers from being directly employed for the photonic device fabrication. To eliminate this loss, we employ a wafer bonding technique to transfer the film to a low-index insulator-on-Si substrate.

\begin{figure}[t]
    \centering
    \includegraphics[width=\linewidth]{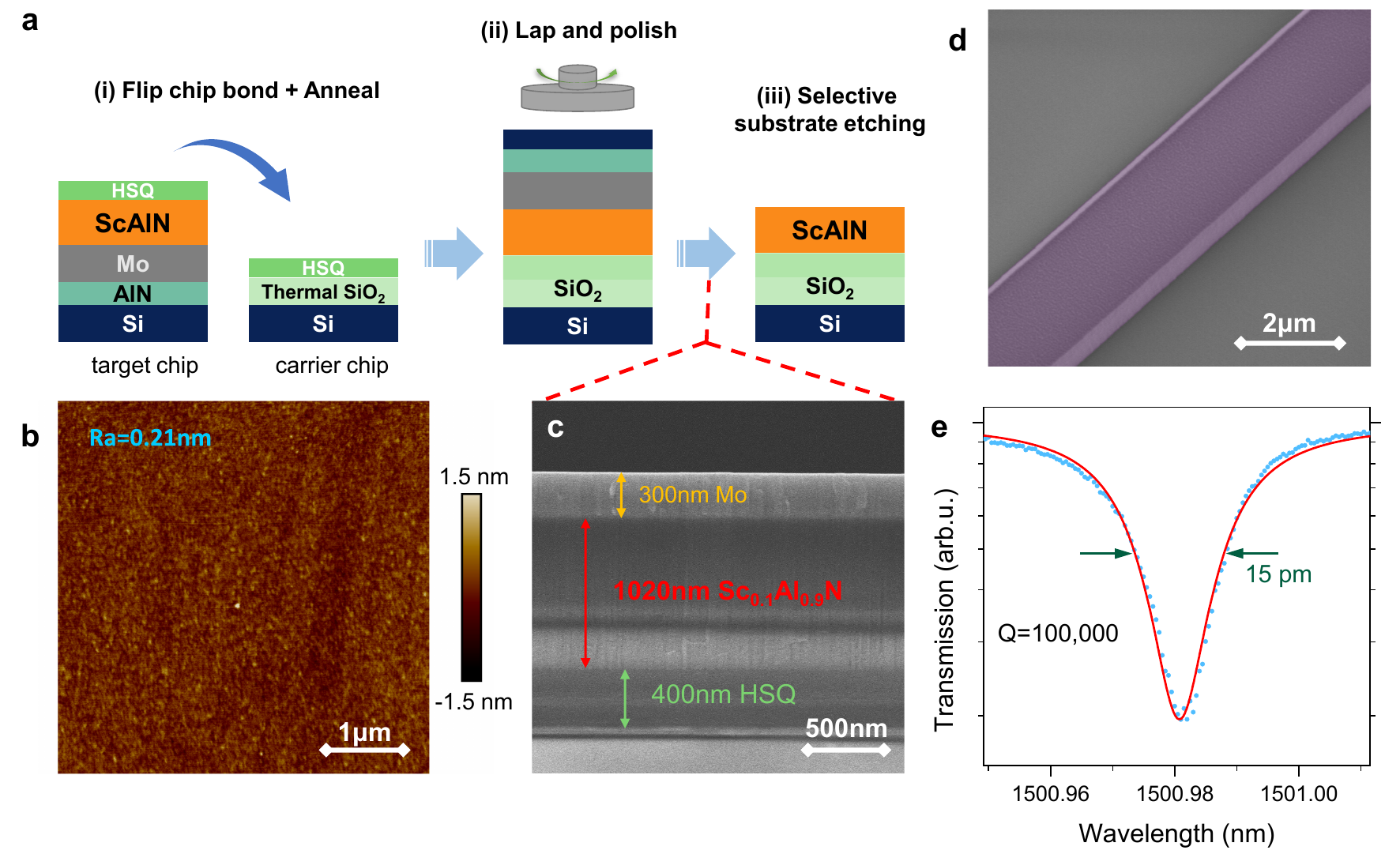}
    \caption{Fabrication and characterization of ScAlN photonics. (a) Preparation of ScAlN-on-insulator platform. HSQ-assisted flip-chip bonding is employed to transfer ScAlN thin film from a metallic substrate onto an insulator-on-Si carrier wafer. The growth substrate is subsequently removed through lapping, polishing, and selective etching procedures. (b) AFM scanning of the surface morphology of polished ScAlN surface, indicating very low surface roughness ($R_a=0.21$\,nm). (c) Cross-sectional SEM image of bonded ScAlN-on-insulator substrate. The Si and AlN buffer layers of the original substrate have been selectively removed, leaving a thin layer of Mo atop ScAlN thin film. No voids are present between the spin-coated HSQ layers, indicating excellent bonding quality. (d) False-color SEM image showing a fraction of an uncoated ScAlN ring resonator. A smooth sidewall is yielded with our dedicated etching recipe. (e) Transmission spectrum of a typical Sc$_{0.1}$Al$_{0.9}$N ring resonator. A linewidth of 15\,pm is observed, which corresponds to a loaded quality factor of $Q_L=1.0\times10^5$.}
    \label{fig2}
\end{figure}

As illustrated in Fig.\,\ref{fig2}a, the bonding process employs spin-coated hydrogen silsesquioxane (HSQ)\cite{2009-hsqbond,2019-hsqbond} on both the ScAlN donor substrate and the oxide-on-silicon host substrate to provide strong adhesion for bonding.
A mechanical polishing step is employed for ScAlN prior to HSQ spinning to prevent surface grain formation. 
The polished surface morphology is characterized with atomic force microscopy (AFM) (Fig.\,\ref{fig2}b), indicating surface roughness as low as $R_a=0.21$\,nm. After the chips are bonded with the aid of HSQ, a vacuum hot-press system is employed to further enhance the bonding strength. To remove the growth substrate of ScAlN, a boron carbide (B$_4$C) slurry is used for fast mechanical lapping of the Si layer, followed by fine polishing with 2.5\,$\upmu$m diamond slurry. 
The remaining layers are subsequently removed with selective etching recipes, resulting in a clean ScAlN surface on a metal-free insulator substrate. Fig.\,\ref{fig2}c shows a cross-sectional SEM image of the bonded chip, revealing smooth and void-free layer boundaries, indicating excellent bonding quality. 

A reactive ion etching (RIE) recipe, which involves a gas mixture comprising Cl$_2$, BCl$_3$, Ar, and N$_2$, is meticulously optimized for etching ScAlN, yielding an etch rate of about 81\,nm/min and a selectivity of Sc$_{0.1}$Al$_{0.9}$N:SiO$_2\approx1:1$. 
Integrated photonic devices are subsequently fabricated, including microring resonators, and grating couplers (GCs) that facilitate optical interfacing with waveguides\cite{2012-chi-aln,2024-sihao-scalnring}. A false-color SEM image showing the fabricated waveguide is shown in Fig.\,\ref{fig2}d. Measurement of the devices is carried out with a tunable laser source (Santec 710) using an array of cleaved single-mode polarization-maintaining fibers aligned to the input/output GCs. We observe a typical peak coupling efficiency per GC of approximately 10\%. The transmission spectrum for an under-coupled Sc$_{0.1}$Al$_{0.9}$N ring resonator is plotted in Fig.\,\ref{fig2}e. From this, we extract an intrinsic quality factor $Q_\mathrm{in}$ of roughly $1.4\times10^5$. In addition, the Sc$_{0.3}$Al$_{0.7}$N ring resonators demonstrate a $Q_\mathrm{in}$ of approximately $3.6\times10^4$. Both results are consistent with recent work in the literature\cite{2024-sihao-scalnring,2024-frideman-scalnring}. The increased loss observed with higher Sc concentrations could be attributed to a greater likelihood of forming rock-salt phase ScAlN during deposition, which possesses an exceptionally narrow bandgap of around 1.5\,eV, resulting in additional undesired absorption\cite{2015-deng-ScNbandgap}. This loss could potentially be mitigated by employing single-crystal ScAlN films.

\subsection*{Pockels effect of ScAlN: Measurements and first-principles studies}\label{sec4}
The Pockels coefficients of ScAlN can be well-characterized and studied by the realization of low-loss microring resonators. Since our film's $c$-axis is oriented perpendicular to the plane ($z$-direction), the refractive index change due to the presence of an applied electrical field can be expressed as $\Delta\left(\frac{1}{n^2}\right)_{ij}=\Sigma_{k=\{x,y,z\}}r_{ij}E_k$, where $r_{ij}$ is the $6\times3$ EO coefficient matrix, $E_x, E_y$ are the in-plane components of the applied RF field, and $E_z$ is the out-of-plane component. Using this definition, we find that the refractive indices for the propagating modes in both TE ($n_{x,y}$) and TM ($n_z$) polarizations can be directly altered with the presence of $E_z$, which leads to a corresponding shift in the microring's resonance frequency. On the other hand, if $E_{x,y}$ is applied, the index ellipsoid will only shift along rotated principal axes, while $n_{x,y}$ and $n_z$ remain unchanged. It is worth noting that rotations of the crystallographic orientation in the $x$-$y$ plane do not alter the effective EO tensor formulation, which suggests that the effective EO response remains valid in our polycrystalline films. 

\begin{figure}[t]
    \centering
    \includegraphics[width=\textwidth]{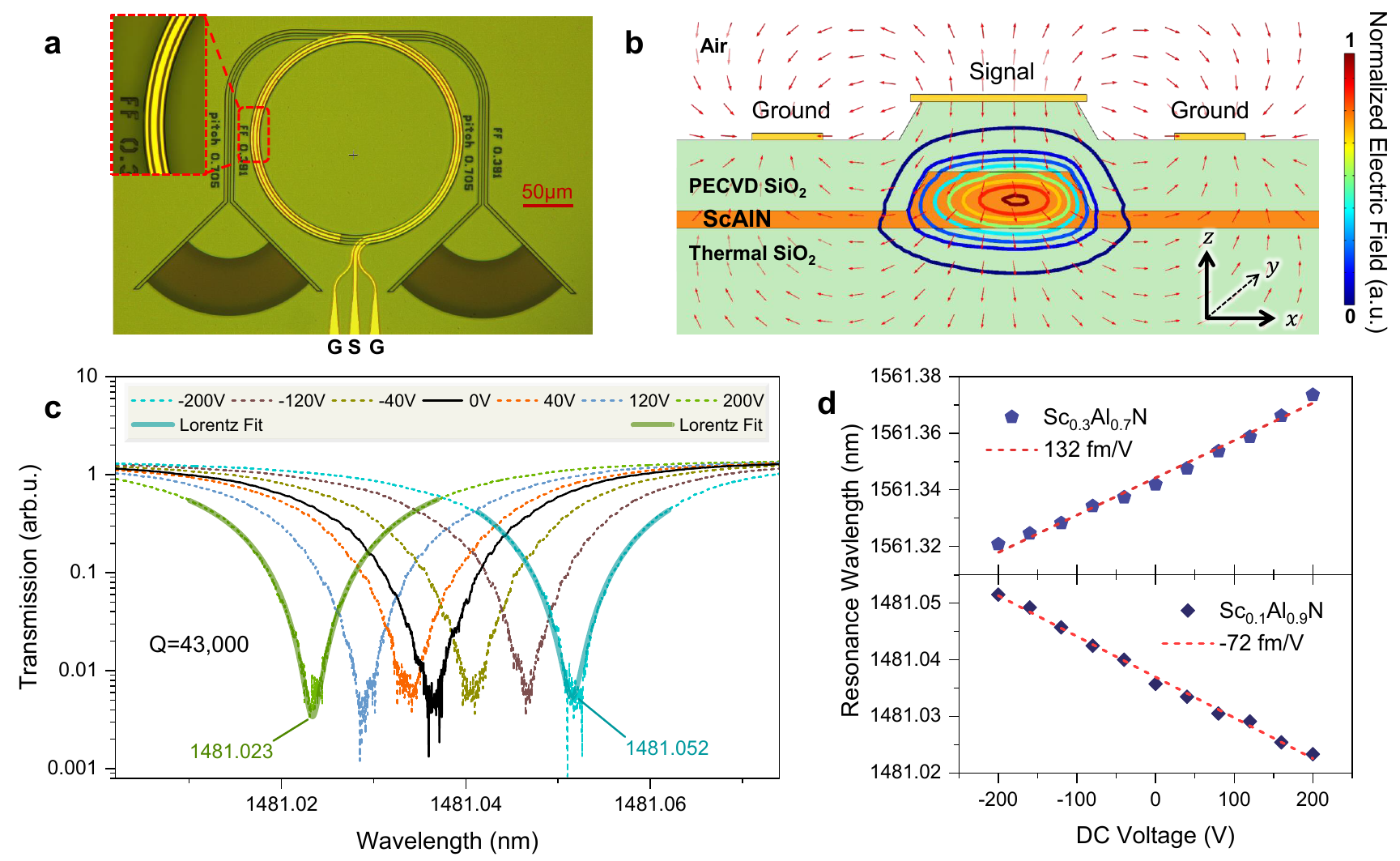}
    \caption{Experimental measurement of ScAlN's Pockels coefficients. (a) Microscope image of a Sc$_{0.1}$Al$_{0.9}$N EOM device. Inset: Zoomed-in view of the GSG electrodes illustrating good alignment with the buried ring resonator. (b) Cross-sectional FDTD simulation of the EOM device. The colored contour lines depict the normalized optical mode profile. The red arrows represent the static electric field when voltage is applied on the signal electrode (length of arrows not to scale). (c) Transmission spectra of a near-critically coupled TE-mode resonance under different applied voltages. Lorentzian fits for $\pm$200\,V are shown in bold. (d) Relationship of TE-mode resonance wavelength with applied voltage. A linear tuning of $-72$\,fm/V and $+132$\,fm/V is observed for Sc$_{0.1}$Al$_{0.9}$N and Sc$_{0.3}$Al$_{0.7}$N, respectively. }
    \label{fig3}
\end{figure}

To generate a vertical electric field ($E_z$) that effectively couples with the optical mode in the ScAlN microring, we employ co-planar ground–signal–ground (GSG) electrodes on top of the ScAlN layer.  Given that electrodes are typically composed of lossy metallic materials, we also design a 1$\upmu$m-thick SiO$_2$ cladding as a buffer layer. The electrode pattern is created using a PMMA e-beam lithography technique, followed by electron beam evaporation of Cr/Au layers (thickness of 10/120 nm respectively) over the cladding and subsequent lift-off process.  Alignment of the electrodes with the ring resonator is achieved using pre-made metallic markers on the chip during e-beam exposures. Fig.\,\ref{fig3}a presents a microscope image of a fabricated electrically-tunable Sc$_{0.1}$Al$_{0.9}$N microring resonator, with the inset offering a zoomed-in view of the precise electrode-ring alignment. The device is consisted of input/output GCs and a ring resonator which is 100\,\textmu m in radius and 2\,\textmu m in width.

The EO modulation performance of the device is characterized by measuring the resonance shift under varying DC voltages. A set of GSG probes is used to provide electrical connection to the electrodes. The DC voltage is precisely controlled by a Keithley 2450 source measure unit (SMU) and is further amplified by a 20$\times$ linear voltage amplifier. As depicted in Fig.\,\ref{fig3}c, a near-critically coupled TE-mode resonance around 1481\,nm is observed in a Sc$_{0.1}$Al$_{0.9}$N device, with an extinction ratio exceeding 22\,dB and a quality factor of around $4.3\times10^4$. A range of DC voltages from -200\,V to 200\,V with increments of 40\,V are applied to the device, and the optical transmission spectrum at each voltage step is measured . A Lorentzian fit is performed for each spectrum to determine the resonance wavelength, which exhibits a linear shift of $-72$\,fm/V in response to the applied voltage as illustrated in Fig.\,\ref{fig3}d.  For Sc$_{0.3}$Al$_{0.7}$N devices, we instead observe a linear tuning rate of $+132$\,fm per volt. 

The $r_{13}$ Pockels coefficient can be derived from the slope of this linear relationship as
\begin{equation}\label{eqn:PockelsCalc}
    r_{13}\approx\frac{-2}{n_o^2\lambda_0\eta}\frac{\Delta\lambda/U}{
E_{z,\mathrm{avg}}/U}.
\end{equation}
Here, $\lambda_0$ represents the resonance wavelength in the absence of an electric field, $\eta$ the electrode coverage ratio over the ring resonator, and $E_{z,\mathrm{avg}}$ the effective electric field averaged based on its overlap with the optical mode. The term $\Delta\lambda/U$ is directly obtained from the linear fit’s slope. However, $E_{z,\mathrm{avg}}/U$ cannot be measured directly, thus requiring us to employ finite-difference time-domain (FDTD) simulations to estimate its magnitude. The simulation is carried out in COMSOL Multiphysics by incorporating the optical mode profile and the static electric field distribution with $U=1$\,V applied to the signal electrode (Fig.\,\ref{fig3}b). Accurate modeling requires knowledge of the optical refractive indices and the permittivities of ScAlN. We experimentally measured the refractive indices using optical ellipsometry (see SI Fig.\,S5). As for static-field permittivity, it was assigned to be $11$ for Sc$_{0.1}$Al$_{0.9}$N and $25.5$ for Sc$_{0.3}$Al$_{0.7}$N, based on findings from previous research\cite{2010-gunilla-scalnepsilon,2022-joseph-scalnepsilon,2022-shao-scalnepsilon30}. 
In addition, an analogous experiment is carried out on single crystalline AlN, serving as a data-point for 0\% Sc. 

\begin{figure}[t]
    \centering
    \includegraphics[width=\linewidth]{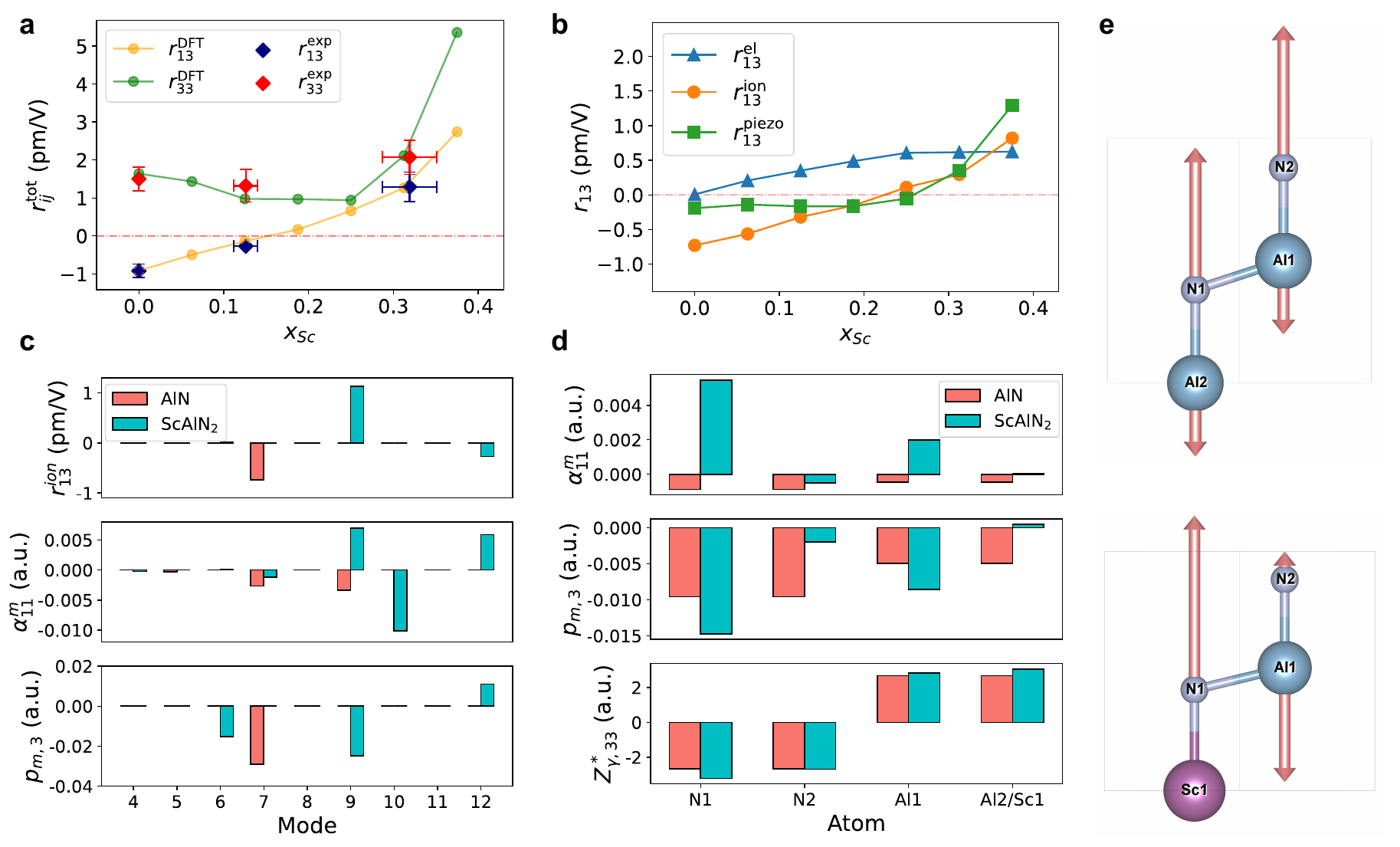}
    \caption{First-principles studies of the EO coefficients of ScAlN. (a) Comparison between experiment and calculated total EO coefficient $r_{13}$ and $r_{33}$. Error bars represent estimated 2$\sigma$ uncertainties. (b) Separated electronic, ionic and piezoelectric contributions to $r_{13}$ coefficient. (c) Mode-by-mode decomposition of $r_{13}^{\mathrm{ion}}$ in AlN and $\mathrm{Sc_{0.5}Al_{0.5}N}$, with their corresponding vibrational eigenmodes depicted in (e). Phonon modes are labeled in ascending order of frequency. $\alpha_{11}^m$ and $p_{m,3}$ are components of the Raman susceptibility and mode polarity induced by phonon mode $m$. (d) Atomic contributions to $\alpha_{11}^m$, $p_{m,3}$ and Born effective charge $Z^*_{\gamma,33}$ for the dominant phonon modes ($\mathrm{A}_1$(TO) modes in AlN and ScAlN$_2$). (e) Atomic vibrations of $\mathrm{A}_1$(TO) modes in AlN (625 $\mathrm{cm}^{-1}$) and in ScAlN$_2$ (540 $\mathrm{cm}^{-1}$). Eigenvectors of phonon modes are represented by arrows.}
    \label{fig4}
\end{figure}

From these experiments, we find the $r_{13}$ coefficient for Sc$_x$Al$_{1-x}$N to be $-0.92\pm0.18, -0.27\pm0.09$ and $1.29\pm0.39$\,pm/V for $x=0, 0.1$, and $0.3$, respectively. The $r_{33}$ coefficients are determined to be $1.50\pm0.31, 1.32\pm0.43$ and $2.07\pm0.45$\,pm/V. The estimated errors account for a potential deviation of $2\sigma=20\%$ between the FDTD model and the actual experimental performance in addition to other measurement inaccuracies. 
All these results are consistent with our calculations, as illustrated in Fig.\,\ref{fig4}a. We note that the discrepancy between the measured and calculated $r_{33}$ values may be ascribed to its sensitivity to stress in the film and actual atomic arrangement in the lattice \cite{Haochen_EOpaper}.

Intriguingly, our observations reveal that $r_{13}$ coefficient undergoes a transition from negative to positive values at 16\% Sc concentration, as shown in Fig.\,\ref{fig4}a.
To understand this phenomenon, we individually analyze the electronic, ionic and piezoelectric contributions to $r_{13}$.
Fig.\,\ref{fig4}b illustrates that the piezoelectric component $r_{13}^\mathrm{piezo}$ is close to zero for Sc concentrations below 25\%.
The electronic contribution $r_{13}^\mathrm{el}$ is always positive and incrementally rises with increasing Sc content.
The overall sign change is due to the ionic response $r_{13}^\mathrm{ion}$, which is negative up to 22\% Sc and then turns positive. 

To clarify the sign change in $r_{13}^\mathrm{ion}$, we investigate pure AlN (0\% Sc) and $\mathrm{ScAlN_2}$ (50\% Sc), as depicted in Fig.\,\ref{fig4}e.
The Sc concentration range between these two structures encompasses the sign-flipping point at 22\%.
Both structures are modeled in 4-atom primitive cells that contain only 12 phonon modes, simplifying the analysis.
In Fig.\,\ref{fig4}c, we perform mode-by-mode decomposition of $r_{13}^\mathrm{ion}$ in the two structures, with nine optical phonon modes (4-12) sequentially indexed in ascending order of phonon frequency. 
Our findings indicate that in each structure one mode dominates the ionic EO contribution: mode 7 in AlN and mode 9 in $\mathrm{ScAlN}_2$, which correspond to $\mathrm{A}_1$(TO) modes in both cases; the atomic displacements that characterize these modes are depicted in Fig.~\ref{fig4}e. 

Comparing the Raman susceptibility $\alpha_{11}^m$ and the mode polarity $p_{m,3}$ of  two $\mathrm{A}_1$ modes, the sign difference of $r_{33}^{\mathrm{ion}}$ between AlN and $\mathrm{ScAlN}_2$ originates from the sign flipping of the Raman susceptibility.
Given that the eigendisplacements of these two modes are in the $c$ direction, we can simplify Eq.\,\ref{raman_polarity} (see the Methods section) to $\alpha_{1 1}^m =\sqrt{\Omega_0} \sum_{\gamma=1}^{4} \frac{\partial \chi_{11}^{(1)}}{\partial \tau_{\gamma,3}} u_m(\gamma,3)$ and $p_{m, 3} =\sum_{\gamma=1}^{4} Z_{\gamma, 33}^* u_m(\gamma,3)$, where $\gamma$ labels the atoms within the cell.
This enables us to analyze the contributions from individual atoms to $\alpha_{1 1}^m$ and $p_{m, 3}$ in $\mathrm{A}_1$ modes of AlN and $\mathrm{ScAlN}_2$, as shown in Fig.\,\ref{fig4}d.
%Since $Z_{\gamma, 33}^*$ in both modes are close, the disparity in mode polarity $p_{m,3}$ is solely attributable to $u_m(\gamma,3)$. 
Relative to AlN, $\mathrm{ScAlN_2}$ exhibits a significant enhancement in mode polarities contributed by N1 and Al1, while the contributions from N2 and Al2/Sc1 are diminished.
Additionally, the Raman susceptibilities $\alpha_{11}^m$ ascribed to N1 and Al1 are significantly enhanced (up to sixfold) and have flipped sign in $\mathrm{ScAlN_2}$%, leading to the positive $r_{13}^{\mathrm{ion}}$ in mode 9.

%From the analyses above, the alloyed Sc atom modifies the ionic EO coefficient $r_{13}^{\mathrm{ion}}$ by amplifying the displacement of adjacent N atoms along $c$ direction and the nearest Al atoms out of plane, which possess a positive Raman susceptibility. 
From the analyses above, the alloyed Sc atom modifies the ionic EO coefficient $r_{13}^{\mathrm{ion}}$ by amplifying the displacement along the $c$ direction of adjacent N and Al atoms, which possess a positive Raman susceptibility.
This counteracts the negative Raman susceptibility contributed by phonon modes similar to {\color{red} $\mathrm{A_1}$} in AlN.
Combined with the electronic and piezoelectric contributions, this leads to the observed sign change in total $r_{13}$ around 16\%.

\section*{Discussion}
%\question{talk about theory first} 
In summary, we performed a comprehensive investigation into the ferroelectric nitride ScAlN and its electro-optic properties. Theoretical predictions based on first-principles calculations suggest that increased Sc concentrations hold significant potential for enhancing ScAlN’s electro-optic performance. Experimentally, our research has successfully developed a robust flip-chip bonding technique, which enables the production of ScAlN-on-insulator substrates with low optical loss, and drives progress in photonic device fabrication with thin-film ScAlN. Moreover, our experimental determination of the Pockels coefficients for Sc$_x$Al$_{1-x}$N, with Sc concentrations of $x=0, 0.1,$ and $0.3$, further corroborates our theoretical models. 

Despite current numbers falling behind expectations, several roads exist towards further exploiting ScAlN for enhanced EO effects. Studies have demonstrated that the use of multiple quantum wells can significantly enhance the Pockels effect in III-N semiconductors\cite{2015-Wei-AlGaNMWQ,2023-zetian-AlGaNMQW}. Meanwhile, global efforts in material science are also paving the way towards the synthesis of optical-quality ScAlN films at high Sc concentrations of $x>0.4$\cite{2018-yuanLu-41percpiezo,2019-simon-ferroscaln,2021-ambacher-scalnproperty}, which should intrinsically possess larger Pockels coefficients. 

Overall, the ScAlN-on-insulator platform we have developed not only facilitates further exploration into the electro-optic properties of ScAlN, but also establishes it as a promising candidate for diverse photonic applications, such as second-harmonic generation\cite{2023-domaincontrol}, acousto-optic devices\cite{2024-kewei-acoustoptic}, and integration with on-chip lasers\cite{2019-Tang-III_V_laser}. While current ScAlN-based photonic devices' performance could be limited by larger optical loss compared to that of SiN and LN, future material deposition may be optimized to reduce intrinsic absorption loss, and scattering loss could be mitigated by optimizing etching recipes or adopting rib-loaded hybrid configurations \cite{yoshioka2024cmoscompatible}. Along with its CMOS-compatibility and exceptional mechanical properties, ScAlN stands poised to herald a new epoch of versatile semiconductor platforms.

\emph{Note added:} An arXiv version of this manuscript is available online\cite{yang2024unveiling}. As we were preparing to submit this manuscript, we became aware of concurrent postings \cite{yoshioka2024cmoscompatiblearxiv,xu2024siintegratedscaln} that also investigated the electro-optic effect of ScAlN.

\section*{Methods}\label{sec:methods}
\subsection*{XPS and XRD characterization}
XPS measurements were performed using a PHI VersaProbe II spectrometer using a monochromatic 1486.7\,eV Al K$\alpha$ X-ray source with a 0.47\,eV system resolution. The energy scale was calibrated using Cu 2$p$\,3/2 (932.67\,eV) and Au 4$f$\,7/2 (84.00\,eV) peaks on a clean copper plate and a clean gold foil. All measurements were collected with a pass energy of 27\,eV and a step size of 0.1\,eV. To remove surface oxidation, prior to measurements, the surface was cleaned using 3\,mm$\times$3\,mm rasterized Ar-ion sputtering operating at 4\,kV and 161\,nA for 5 minutes, corresponding to a precalibrated sputtering depth of 54\,nm on SiO$_2$.  An electron flood gun was used to compensate for charging effects.

Structural characterization was performed using a Rigaku SmartLab X-ray diffractometer equipped with a Cu K$\alpha$ source with a beam energy of 8.04\,keV ($\lambda=1.5406$\AA). Out-of-plane XRD measurements were performed using a Bragg-Brentano geometry.  Rocking curve measurements and in-plane XRD were performed using a parallel beam geometry.  The incidence angle for in-plane measurements was $\omega = 0.3^{\circ}$. The results of in-plane XRD are summarized in SI section I.

\subsection*{Density Functional Perturbation Theory}
In the framework of the density functional perturbation theory \cite{veithen2005nonlinear}, EO coefficients $r_{ijk}$ can be written as the sum of electronic, piezoelectric, and ionic contributions:
\begin{equation}
    \label{EO_three}
    r_{i j k}=r_{i j k}^{\mathrm{el}}+r_{i j k}^{\mathrm{piezo}}+r_{i j k}^{\mathrm{ion}} =\frac{-8 \pi}{n_i^2 n_j^2} \chi_{i j k}^{(2)}+\sum_{\mu, \nu=1}^3 p_{i j \mu \nu} d_{k \mu \nu}+\frac{-4 \pi}{n_i^2 n_j^2 \sqrt{\Omega_0}} \sum_m \frac{\alpha_{i j}^m p_k^m}{\omega_m^2}.
\end{equation}
The electrical term $\chi^{(2)}_{ijk}$ is the nonlinear optical susceptibility. 
In the piezoelectric term, $d_{k \mu \nu}$ is the piezoelectric strain coefficient and $p_{i j \mu \nu}$ is the elasto-optic coefficient.
In the ionic term, $\Omega_0$ is the volume of the cell, and $\omega_m$ is the phonon frequency of mode $m$.
The Raman susceptibility $\alpha^m_{ij}$ of mode $m$ and the mode polarity $p_{m,k}$ along direction $k$ are defined as:
\begin{equation}
    \label{raman_polarity}
    \alpha_{i j}^m=\sqrt{\Omega_0} \sum_{\kappa, \beta} \frac{\partial \chi_{i j}^{(1)}}{\partial \tau_{\kappa, \beta}} u_m(\kappa \beta), \quad
    p_{m, k}=\sum_{\kappa, \beta} Z_{\kappa, k \beta}^* u_m(\kappa \beta),
\end{equation}
where $u_m(\kappa \beta)$ is the phonon eigendisplacement of atom $\kappa$ along direction $\beta$ in mode $m$,
$\chi^{(1)}_{ij}$ is the linear electric susceptibility, $\tau_{\kappa,\beta}$ represents the atomic displacement of atom $\kappa$ along direction $\beta$, and
$Z^*_{\kappa,k \beta}$ is the $k$-component of Born effective charge of atom $\kappa$ moving along direction $\beta$. 

\subsection*{Polishing of ScAlN thin film}
The as-grown ScAlN films have a significant root-mean-squared (RMS) surface roughness ($R_a$) exceeding 4\,nm. Such roughness increases with Sc concentration, adversely affects bonding quality, and introduces additional scattering loss in photonic devices. Therefore, it is critical to polish the ScAlN thin film prior to the device processing. Widely used techniques, such as dry etching and chemical-mechanical polishing (CMP), have been discussed in Ref. \cite{2024-sihao-scalnring}. However, these methods have not yet met the demand of photonic platforms which typically require an $R_a$ of less than 0.2\,nm to minimize scattering loss. Moreover, a substantial amount of material was removed with these two methods. The limited selectivity of both dry etching, and CMP which employ alkaline-based silica slurry, may contribute to these issues, as they do not differentiate effectively between peaks and valleys on the surface while possessing a high material removal rate (MRR). In contrast, pure mechanical polishing using an oil-based diamond slurry achieves a moderate MRR and exhibits remarkable selectivity for surface protrusions. Our experiment utilizes a soft polishing pad (POLITEX REG II) in conjunction with an oil-based 0.25um diamond slurry (SPLENDIS-ECO O/M 0.25$\_$1). The polishing parameters are as follows: table speed at 40\,rpm, spindle speed at 25\,rpm, slurry flow rate at 6\,mL/min, and head load at 100\,g/cm$^2$. The MRR achieved is approximately 6.5\,nm/min. A brief polishing time of 5 to 10 minutes is sufficient to reduce the $R_a$ to approximately 0.2\,nm, demonstrating significantly greater efficiency than both dry etching and CMP. A conventional cleaning protocol is employed to eliminate residual diamond particles, utilizing acetone and IPA sonication followed by careful wiping with cotton swabs.

\subsection*{HSQ-assisted flip-chip bonding}
Flip-chip bonding plays a key role in preparing ScAlN-on-insulator films for the subsequent fabrication of photonic devices. This process necessitates a void-free bonding interface and the preservation of film uniformity during the removal of the original substrate. To mitigate the impact of surface grains on bonding quality, we commence with the polishing of the as-grown ScAlN surface. The polished ScAlN chip (target) and a 6.0\,$\upmu$m-thick thermal SiO$_2$-on-Si chip (carrier) undergo a comprehensive cleaning process using acetone and IPA sonication, followed by O$_{2}$ plasma activation. HSQ is spin-coated onto both substrates at 3000\,rpm and subject to a hard-bake at 180$\degree$C to ensure complete solvent evaporation. Subsequently, the target is flipped and bonded to the carrier at room temperature applying a force of 2.5kg in a Tresky T-3002-PRO chip bonder. To reinforce the bond strength, the chip assembly is annealed at 400$\degree$C under a 50kg force for 10 hours within a vacuum hot-press system.

To remove the original ScAlN growth substrate, boron carbide (B$_4$C) slurry is used for rapid lapping of Si, followed by polishing with 2.5\,$\upmu$m diamond slurry to alleviate deep scratches and minimize non-uniformity due to edge-rounding. The residual Si layer, with a thickness ranging from 50-80\,$\upmu$m, is selectively dry-etched using SF$_6$ plasma, which stops at the AlN layer. To etch through the AlN, we utilize a Cl$_2$/BCl$_3$ based RIE etch. Finally, an RCA1 solution (NH$_4$OH:H$_2$O$_2$:H$_2$O=1:1:4) is employed to eliminate all traces of Mo, yielding a clear ScAlN surface. Although RCA1 is a basic solution that typically reacts with Al$^{3+}$, its rapid Mo etching rate of over 300\,nm/min ensures minimal ScAlN exposure time, thereby preserving the surface morphology and the performance of the subsequent devices.

\subsection*{Photonic waveguide fabrication}
Given the low etching selectivity of ScAlN, the use of a robust hard mask is essential. Thick HSQ resist has been demonstrated feasible for this goal\cite{2024-sihao-scalnring}, but the wet processes involving alkaline or acidic solutions pose a risk of impairing ScAlN's optical properties. By contrast, we have devised an improved process flow as follows: 400\,nm thick PECVD silicon nitride (SiN) is deposited as a hard mask, and patterns are defined using ebeam lithography with CSAR 62 resist.  The patterns are then transferred to the SiN layer via a CHF$_3$/O$_2$-based RIE process, after which the resist is dissolved using hot NMP, acetone, and IPA. The ScAlN is subsequently dry etched following our optimized recipe (Cl$_2$/BCl$_3$/Ar/N$_2$=10/3/3/4\,sccm, DC=120\,W, ICP=800\,W, pressure 4\,mTorr). The remaining SiN etch mask is removed with RIE, which is highly selective against ScAlN. This method yields a smooth waveguide sidewall and high-resolution lithography. As a final step, a 1$\upmu$m-thick PECVD SiO$_2$ cladding is deposited on the patterned device, facilitating subsequent electrode fabrication.

\subsubsection*{Extracting Pockels coefficients from measured resonance shift}
By expanding the individual matrix elements, we find the effective index change for $n_{x,y}$ and $n_z$ to be
\begin{align}\label{eqn:indexchange}
    \begin{split}
        n_{x,y}&=n_o-\frac{1}{2}r_{13}n_o^3E_z,\\
        n_z&=n_e-\frac{1}{2}r_{33}n_e^3E_z,
    \end{split}
\end{align}
where $n_o$ and $n_e$ are the ordinary and extraordinary refractive indices respectively. 
Here we consider a case where $E_z$ is applied and TE mode propagates in the ring resonator. The relative refractive index change in the lateral direction under the influence of $E_z$ is described by, 
\begin{equation}
    \frac{\Delta n_{x,y}}{n_o}=-\frac{1}{2}r_{13}n_o^2E_z.
\end{equation}
The resonance condition of the microring satisfies 
\begin{equation}
    n_\mathrm{eff}L=m\lambda,
\end{equation}
where $n_\mathrm{eff}$ represents the effective refractive index of the optical mode, $L$ the round-trip length ($2\pi r$ in a circular ring), $m$ the integer order of resonance, and $\lambda$ the resonance wavelength. With the assumption that the relative change in effective mode index is the same as that of the material's refractive index, we have 
\begin{equation}\label{Eqn:BarePockelsCalc}
    \frac{\Delta n_{x,y}}{n_o}=\frac{\Delta n_\mathrm{eff}}{n_{0,\mathrm{eff}}}=\frac{\Delta\lambda}{\lambda_0}=-\frac{1}{2}r_{13}n_o^2E_z.
\end{equation}
Since both the electric field and the mode distribution are non-uniform, an averaged electric field needs to be used to reflect upon the effective EO coupling. The semi-classical EO interaction Hamiltonian is expressed as
\begin{equation}
    H=-\hbar gEa^\dagger a,
\end{equation}
where $a$ and $a^\dagger$ are the optical annihilation and creation operators, and $g$ denoting certain coupling coefficient. It is thus straightforward to show that the effective electric field should be normalized to the intensity of the optical field, 
\begin{equation}
    E_{z,\mathrm{avg}}=\iint_A \varepsilon_k E_{\mathrm{op},k}^2E_z\mathrm{d}S \left/ \iint_\infty \bm{\upvarepsilon}\cdot \bm{\mathrm{E}}_{\mathrm{op}}^2\mathrm{d}S\right. ,
\end{equation}
where $k\in\{x,y\}$, and $A$ denotes the cross-section of the ScAlN waveguide. Substituting this expression into Eqn. (\ref{Eqn:BarePockelsCalc}), we can arrive at Eqn. (\ref{eqn:PockelsCalc}) for the calculation of Pockels coefficients.

\vspace{2 mm}
\section*{Data availability} The data that support the findings of this study are available from the corresponding authors upon reasonable request.

\newpage

\section*{Acknowledgments} 
This work is supported by SUPREME, one of seven centers in JUMP 2.0, a Semiconductor Research Corporation (SRC) program co-sponsored by DARPA. HXT acknowledges support from AFOSR through grant number FA9550-23-1-0338 and from NSF through award number 2235377.  HW was also supported by The Army Research Office under grant number W911NF-22-1-0139. Calculations in this work used Bridges-2 at Pittsburgh Supercomputing Center (PSC) through allocation DMR070069 from the Advanced Cyberinfrastructure Coordination Ecosystem: Services \& Support (ACCESS) program, which is supported by NSF grant No. 2138259, 2138286, 2138307, 2137603 and 2138296.
This research also used resources of the National Energy Research Scientific Computing Center, a DOE Office of Science User Facility, under Contract No. DE-AC02-05CH11231 using NERSC award BES-ERCAP0028497. The facilities used for device fabrication and characterization were supported by the Yale SEAS Cleanroom, the Yale Institute for Nanoscience and Quantum Engineering (YINQE), and the Yale West Campus Materials Characterization Core. The authors would like to express their gratitude to F. Rana and D. Jena for valuable discussions,  to Dr. Yong Sun, Dr. Michael Rooks, Dr. Lauren McCabe, Dr. Yeongjae Shin, and Kelly Woods for assistance in device fabrication, and to Dr. Min Li for his support on material characterization. 

\section*{Author Contributions Statement}
G.Y., H.W., C.G.V., and H.T. conceived the research. G.Y. and H.X. performed the experiments. H.W., S.M., and C.G.V. conducted the density functional theory calculations. T.W. and C.H. characterized the material properties. G.Y., H.X., C.H. and M.S. developed the theory and performed the simulations regarding electro-optic interactions. M.L, C.G.V., and H.T. supervised the project. G.Y., H.W. and T.W wrote the manuscript with input from all authors.

\section*{Competing Interests Statement}
The authors declare no competing interests.

\section*{Figures}

%%===========================================================================================%%



\end{document}